\documentclass[reqno,12pt]{amsart} %maths
\usepackage[top=1.5in,right=1.125in,left=1.125in,bottom=1.5in]{geometry}
\usepackage{amssymb}
\usepackage{color}
\usepackage{tikz}
\usetikzlibrary{positioning,decorations.pathmorphing}
\usetikzlibrary{positioning,decorations.pathmorphing}
\definecolor{red}{rgb}{0.7,0,0}
\definecolor{grey}{RGB}{112,112,112}
\definecolor{blue}{RGB}{034,113,179}
\usepackage[colorlinks=true,citecolor=blue,linkcolor=red,urlcolor=grey,backref=page]{hyperref}
\newcommand{\koniec}{\begin{flushright}  $\Box $ \end{flushright}}
\newtheorem{theo}{Theorem}[section]

\newtheorem{col}[theo]{Corollary}

\theoremstyle{remark}

\newcounter{mnotecount}[section]

\renewcommand{\themnotecount}{\thesection.\arabic{mnotecount}}

\newcommand{\mnote}[1]%{}%
{\protect{\stepcounter{mnotecount}}$^{\mbox{\footnotesize
$%\!\!\!\!\!\!\,
\bullet$\themnotecount}}$ \marginpar{%\color{red}%
\raggedright\tiny\em
$\!\!\!\!\!\!\,\bullet$\themnotecount: #1} }

\newcommand{\C}{\mathbb{C}}
\newcommand{\Z}{\mathbb{Z}}

\newcommand{\R}{\mathbb{R}}

\def\Ve{V_{\mbox{eff}}}
\def\p{\partial}
\def\be{\begin{equation}}

\def\ee{\end{equation}}

\def\bea{\begin{eqnarray}}
\def\eea{\end{eqnarray}}

\numberwithin{equation}{section}
\begin{document} \date{July 15th, 2024}
%%%%%%%%%%%%%%%%%%%%%%%%%%%%%%%%%%%%%%%%
\title{Integrability of quantum dots}
\author{Maciej Dunajski}
\address{Department of Applied Mathematics and Theoretical Physics\\ 
University of Cambridge\\ Wilberforce Road, Cambridge CB3 0WA, UK.}
\email{m.dunajski@damtp.cam.ac.uk}
\author{Andrzej Maciejewski}
\address{Janusz Gil Institute of Astronomy\\
University of Zielona G\'ora\\
Licealna 9, 65-417, Zielona G\'ora\\
 Poland}
\email{a.maciejewski@ia.uz.zgora.pl}
\author{Maria Przybylska}
\address{ Institute of Physics\\ 
University of Zielona G\'ora\\
Licealna 9, 65–417, Zielona G\'ora\\
 Poland
}
\email{maria.przybylska@gmail.com}
\maketitle
\begin{abstract}
We determine the frequency ratios $\tau\equiv
\omega_z/\omega_{\rho}$ for which the Hamiltonian system with a potential
\[
V=\frac{1}{r}+\frac{1}{2}\Big({\omega_{\rho}}^2(x^2+y^2)+{\omega_z}^2 z^2\Big)
\]
is completely integrable. We relate this result to the existence of conformal Killing tensors of the associated Eisenhart metric on $\R^{1, 4}$. Finally
we show that trajectories of a particle moving under the influence of the potential $V$ are not unparametrised geodesics of any Riemannian metric on $\R^3$.
\end{abstract}
\section{Introduction}
Consider a particle with unit mass moving in $\R^3$ under the influence of a potential
\be
\label{V1}
V=\frac{\alpha}{r}+\frac{1}{2}\sum_{a, b=1}^3 \omega_{ab} x^ax^b
\ee
where $\alpha$ is a constant, $\omega=\omega_{ab}$ is a symmetric matrix with
constant coefficients, and $r=\sqrt{(x^1)^2+(x^2)^2+(x^3)^2}$. If
$\omega_{ab}=0$, then $V$ corresponds to the Coulomb--Kepler problem which is
completely integrable. If $\alpha=0$ and $\omega\neq 0$ then $V$ corresponds to
the anisotropic harmonic oscillator. If $\alpha$ is negative, and all
eigenvalues of $\omega$ are negative (positive) and equal, then $V$ is a
Newton--Hooke potential arising in a non--relativistic limit of the
Schwarzschild --de Sitter (Schwarzschild anti-de-Sitter) space--time \cite{D23}.
In the Schwarzschild--de Sitter case there exists an equilibrium where the
attractive Newtonian force balances the cosmological repulsion, but this
equilibrium is unstable. If $\alpha>0$ and all eigenvalues of $\omega$ are
positive then the repulsive Coulomb force becomes small away from the origin
where $r=0$, and the attractive harmonic force grows linearly with the distance
from the origin. In this case the equilibrium is stable, and the particle moving
in the potential $V$ is confined to a finite region. This is the quantum
dot\footnote{The original quantum dot involves two charged particles with
Coulomb repulsion and a confining quadratic potential. Potential (\ref{V1})
arises after a transformation to the centre of mass coordinates.} \cite{Q2}.

The subject of this note is the integrability of the quantum dot potential. If
all eigenvalues of $\omega$ are equal then $V=V(r)$ and there exists three first
integrals in involution. We therefore focus on the axi--symmetric case where two
of the eigenvalues of $\omega$ are equal. Diagonalising $\omega$, scaling the
dependent and independent variables to reabsorb some constants so that
$\alpha=1$ and $\omega=\mbox{diag}(1, 1, \tau^2)$, and using the cylindrical
coordinates $(\phi, \rho, z)$ on $\R^3$ leads to the Lagrangian
\be
\label{lagrangian}
L=T-V, \quad \mbox{where}\quad
T=\frac{1}{2}(\dot{z}^2+\dot{\rho}^2+\rho^2\dot{\phi}^2), \quad
V=\frac{1}{\sqrt{\rho^2+z^2}}+\frac{1}{2}\Big(\rho^2+\tau^2 z^2\Big),
\ee
with $\tau^2=1$ corresponding to the spherically symmetric case.  If $\tau^2\neq
1$, there exist two obvious first integrals: the Hamiltonian $H=T+V$, and the
$z$--component of the angular momentum $h=\rho^2\dot{\phi}$. It is known
\cite{Q2, gary1, gary2} that the Hamilton--Jacobi equation with potential
(\ref{lagrangian}) is separable, and therefore the corresponding Hamilton's
equations are completely integrable, iff $\tau^2$ is equal to $1$ or $4$. If
$\tau^2=4$ the additional quadratic first integral is given by
\be
\label{Q2}
Q_2=z\dot{\rho}^2-\rho\dot{z}\dot{\rho}+\frac{z}{\sqrt{\rho^2+z^2}}-\rho^2 z+\frac{z}{\rho^2}h^2.
\ee
The only other known integrable case corresponds to $\tau^2=1/4$. In this case
the potential is not separable, but there exists a first integral quartic in the
velocities which is in involution with $H$ and $h$. It is given by \cite{Q4,
gary1, gary2}
\begin{eqnarray}
\label{Q4}
Q_4&=&\Big(A+\frac{h^2}{\rho}\Big)^2+\Big(\dot{\rho}+\frac{z}{\rho}\dot{z}\Big)^2h^2+(\rho^2+z^2) h^2, \quad \mbox{where}\\
A&=&\rho \dot{z}^2-z\dot{\rho}\dot{z}+\frac{\rho}{\sqrt{\rho^2+z^2}}-\frac{1}{4}\rho z^2.\nonumber
\end{eqnarray}
One way to find this integral is to first consider a restricted initial data
with $h=0$. This makes the Lagrangian (\ref{lagrangian}) symmetric with respect
to interchanging $z$ with $\rho$ and $\tau^2$ with $\tau^{-2}$ as long as it is
combined with the scaling of the independent variable. This leads to a quadratic
first  integral $A$ in (\ref{Q4}) analogous to (\ref{Q2}). While $A$ is not a
first integral if $h\neq 0$, it can be corrected by terms involving $h$ to yield
a first integral  (\ref{Q4}).

In \S\ref{galois} we shall use the  differential Galois theory and its applications to Hamiltonian systems
developed by Morales-Ruiz and Ramis \cite{MR} to 
demonstrate that these three values of $\tau^2$ exhaust all integrable cases.
\begin{theo}
\label{theo1}
The Hamilton's equations resulting from the potential (\ref{lagrangian}) are completely integrable iff
$\tau^2\in\{1, 4, 1/4\}$.
\end{theo}
 It follows from the work of Eisenhart \cite{Eisenhart} that for a given $\tau\in\R$
any integral curve of the Euler--Lagrange equations
with $L$ given by (\ref{lagrangian}) lifts to a null geodesics of a Lorentzian metric $G_{\tau}$ in $(4+1)$ dimensions
\be
\label{Gtau}
G_{\tau}=2dudt+2\Big(\frac{1}{\sqrt{\rho^2+z^2}}+\frac{1}{2}(\rho^2+\tau^2 z^2) \Big)dt^2-dz^2-d\rho^2-\rho^2 d\phi^2.
\ee
In \S\ref{sec_eisenhart} we shall deduce the following Corollary from Theorem \ref{theo1}
\begin{col}
\label{color} The Eisenhart metric (\ref{Gtau}) admits an irreducible conformal Killing tensor different from itself and Lie--derived by $\p/\p\phi$ iff $\tau^2\in\{1, 4, 1/4\}$.
\end{col}

 Solutions to Euler--Lagrange equation in $\R^3$ define a local path geometry: A four--parameter family of unparametrised curves, one curve through any point and in any direction. In the case of (\ref{lagrangian}) this path geometry can be encoded in a pair of second order ODEs for $\rho=\rho(\phi)$ and 
$z=z(\phi)$. In \S\ref{sec_proj} we 
use the projective invariants found in \cite{DE} to 
show that although this path geometry consists of unparametrised geodesics of an equivalence class of affine connections, none of these connections arise as a Levi--Civita connection of a metric
\begin{theo}
\label{theo2}
The path geometry resulting from (\ref{lagrangian}) is not metrisable.
\end{theo}
\section{Meromorphic first integrals in involution}
\label{galois}
In this Section we shall prove Theorem \ref{theo1}. Our approach is
based on the following result \cite{MR} 
\begin{theo}[Morales-Ramis]
If a complex Hamiltonian system is integrable in
the Arnold--Liouville sense with complex meromorphic first
integrals then the identity component of the differential Galois group of variational equations is Abelian.   
\end{theo}
To apply this Theorem in our setup, we shall first pass to the
Cartesian coordinates\footnote{In our proof we allow for the first integrals which are meromorphic
in $(x_1, x_2, x_3, p_1, p_2, p_3)$ as well as $r$. Otherwise we could not account for the Hamiltonian as a meromorphic first integral. See \cite{combot, mp16} where this extension was applied to other potentials.}
$(x_1, x_2, x_3)=(\rho\sin\phi, \rho\cos\phi, z)$, where
the Euler--Lagrange equations of (\ref{lagrangian}) are
\be
\label{eq:11}
\ddot{x}_1=\frac{x_1}{r^3}-x_1, \quad
\ddot{x}_2=\frac{x_2}{r^3}-x_2, \quad
\ddot{x}_3=\frac{x_3}{r^3}-\tau^2 x_3.
\ee
The first
step will be to pick two solutions to (\ref{eq:11}) and consider their particular linearisations
in the form
\be
\label{lin_eq}
  \frac{d^2 w}{d \zeta^2}=r(\zeta) w,
\ee
where $\zeta=\zeta(t), w=w(\zeta)$, and $r(\zeta)$ is a given rational
function which depends on a chosen solution as well as the parameter $\tau$. We shall then use the Kovacic algorithm
\cite{Kovacic:86::}
  to find the necessary conditions for the identity component
  of the differential Galois group of (\ref{lin_eq}) to be Abelian.
  Two particular solutions to (\ref{eq:11})  will yield two sets of conditions each of which constrains the value of
  the parameter $\tau$ in the potential (\ref{lagrangian}). We will find these conditions to be
  \[
    \tau=\frac{m_1}{2k_1}\quad\mbox{and}\quad \frac{1}{\tau}=
    \frac{m_2}{2k_2}
  \]
  where $m_1, m_2$ are integers, and
  $\{k_1, k_2\}\in\{1, 2, 3, 4, 5, 6\}$.
  This simultaneous system of algebraic equations has
  $37$ solutions for $\tau$, and applying the Kovacic algorithm
  to each solution we can determine whether the resulting
  linear equations (\ref{lin_eq}) have solutions
  which are Liouvillian\footnote{Recall, that a set
    of Liouvullian functions is defined recursively starting
    form elementary functions, their integrals, and
    integrals of the resulting functions. This set is closed
under aritmetic operations, composition of functions as well as
differentiation.}. 
    The existence of only Liouvillian solutions
  is equivalent to the solvability of the differential Galois
  group of (\ref{lin_eq}). This in turn is a necessary condition
  for this group to be Abelian, and the corresponding
  Hamiltonian system to be completely integrable. The
  details are as follows:

\noindent
{\bf Proof of Theorem \ref{theo1}.}
Let us take two solutions to (\ref{eq:11})
\begin{eqnarray}
\label{eq:17}
\Gamma_1&=&\left\{x_2=x_3=p_2=p_3=0,\quad 2e = p_1^2 +x_1^2 +\frac{2}{x_1}\right\}\nonumber\\
\Gamma_2&=&\left\{x_1=x_2=p_1=p_2=0,\quad 2e = p_3^2 +\tau^2 x_3^2 + \frac{2}{x_3}\right\}
\end{eqnarray}
where $(p_1, p_2, p_3)$ are the conjugate momenta, and $e$
is a constant corresponding to the first integral $H$.
We now consider the variational equations (
the linearisation $x_i(t)=x_i+\epsilon X_i(t)$)
of these solutions in turn. For the first solution, one
of the variational equations is
\[
%\label{eq:18}
\ddot X_3 =  \left( \frac{1}{{x_1}^3} -\tau^2    \right) X_3.
\]
This equation is equivalent to
\[
%\label{eq:19}
X_3'' + p(\zeta) X_3' + q(\zeta) X_{3}=0
\]
where $\zeta=x_1(t)$ and
\begin{equation}
\label{eq:20}
p(\zeta) = \frac{\zeta^3-1}{\zeta \left( \zeta^3 - 2 e \zeta +2 \right)}, \quad
q(\zeta) =- \frac{\tau^2\zeta^3-1}{\zeta^2 \left( \zeta^3 - 2 e \zeta +2 \right)}.
\end{equation}
Setting 
\[
X_3=w(\zeta)\exp\left[-\frac{1}{2}\int p(s)\mathrm{d}s\right]
\]
we obtain the normal form (\ref{lin_eq})
\begin{equation}
\label{eq:21}
w''=r_1(\zeta)w,\quad\mbox{where}\quad
r_1(\zeta)= \frac{1}{2}p'(\zeta)+\frac{1}{4}p(\zeta)^2-q(\zeta).
\end{equation}
The explicit form  of $ r_{1}$ is
\[
%\label{eq:22}
r_1(\zeta) =\frac{\left(4 \tau^2-1\right) \zeta^6 -4e\left(2  \tau^2+1\right)\zeta^4+2 \left(4 \tau^2+5\right)
   \zeta^3-3}{4 \zeta^2 \left(\zeta^3-2 e \zeta+2\right)^2}.
\]
If $e$ is real and $e \neq 3/2$ then the polynomial $\zeta^3-2 e \zeta+2$ has three different
roots $(\zeta_1, \zeta_2, \zeta_3)$ different from $\zeta_0=0$.  Thus, $(\zeta_0, \dots, \zeta_3)$
are regular singular points of equation~\eqref{eq:21}.
For all these points the difference of exponent is $\Delta=1/2$. If
$\tau\neq 1/2$ then the infinity is also regular singular point with
the difference of exponent $\Delta_{\infty}= 2 \tau$.  If $\tau=1/2$ then the infinity  is a regular point and we can proceed with our algorithm with $\Delta_\infty=1$.
If the system is integrable, then the identity component of the differential Galois group of~\eqref{eq:21} is Abelian, and all its solution
are Liouvillian. The necessary and sufficient conditions are given by the Kovacic algorithm, see \cite{Kovacic:86::}. As the
equation contains parameters, we are only able
to extract the necessary conditions which are
\begin{equation}
\label{eq:23}
\tau = \frac{m_1}{2k_1}, \quad \mbox{for certain}\quad m_1\in\Z, \qquad k_1\in\{1,\ldots, 6\}.
\end{equation}
If the  equation of the form
(\ref{lin_eq}) has a Liouvillian solution then its
certain symmetric power\footnote{A $k$th symmetric power
  of a linear ODE of the form (\ref{lin_eq}) is a linear
  ODE of order $k+1$ whose linearly independent solutions are monomials ${w_1}^k, {w_1}^{k-1}w_2, \dots, w_1{w_2}^{k-1}, {w_2}^{k}$,
  where $w_1(\zeta)$ and $w_2(\zeta)$ are linearly independent
  solutions to (\ref{lin_eq}).}
has a rational solution.  Possible
degrees of the numerator of this solution can be
determined  by exponents  of singular points and the
necessary condition (\ref{eq:23}) was deduced from this fact.

Repeating  this calculation for the  second solution in
(\ref{eq:17})
we obtain  the following form reduced form of variational equation
\be
\label{eq:25}
w''=r_2(\zeta) w
\ee
where
\[
%\label{eq:26}
r_2(\zeta) = -\frac{4 e \left(\tau^2+2\right) \zeta^4+\tau ^2 \left(\tau ^2-4\right) \zeta^6-2
   \left(5 \tau ^2+4\right) \zeta^3+3}{4 \zeta^2 \left(\tau ^2 \zeta^3-2 e \zeta+2\right)^2}.
\]
If $\tau^2=4$ then the infinity  is a regular point and can proceed with our algorithm with $\Delta_\infty=1$. If
\[
%\label{eq:27}
\tau^2(8e^3-27\tau^2)\neq 0 \quad\mbox{and}\quad
\tau^2\neq 4, 
\]
then this equation is Fuchsian.
It has four singular points in $\C$ :
$\zeta_0=0$ and three roots of polynomial $\tau ^2 \zeta^3-2 e \zeta+2$.
The infinity is also regular singular point. The difference of
exponents at finite points  is $\Delta=1/2$ and we find
$\Delta_{\infty}=2/\tau$.   Hence we have additional necessary conditions for the integrability
\begin{equation}
\label{eq:28}
\frac{1}{\tau} = \frac{m_2}{2k_2}, \quad
\mbox{for certain}\quad m_2\in\Z, \qquad k_2\in\{1,\ldots, 6\}.
\end{equation}
Now, from (\ref{eq:23}) and (\ref{eq:28}) we deduce that
\[
%\label{eq:29}
m_1 m_2 = 4 k_1 k_2 \quad\mbox{for}\quad
k_1, k_2\in \{1,\ldots, 6\}. 
\]
This equation has only a finite number of solutions
$(m_1 , m_2)\in \Z^2$. All these solutions  give 37 possible values  for $\tau>0$, 
namely, $ \tau =s $ or $\tau =1/s $ where
\begin{equation}
\label{eq:30}
s\in \{1, 2, 3, 4, 5, 6, 8, 10, 12\} \cup \{ 6/5, 5/4, 4/3, 3/2, 8/5, 5/3,  12/5, 5/2, 8/3,  10/3\}. 
\end{equation}
For each of these values of $\tau$ we can check, using the Kovacic algorithm,
whether both variational equations have  a Liouvillian solution. If one of them
does not admit such solutions, then the identity component of the differential
Galois group of this equation is not solvable. Therefore, it is also
non--Abelian, and the system is not integrable. In this way we exclude all the
cases but $\tau^{2}\in\{1/4,1, 4\}$.

Conversely, if $\tau^2=1$ then the potential is radially symmetric, so the
system is integrable. If $\tau^2=4$ or $1/4$ then it is also integrable with the
additional first integral given by (\ref{Q2}) or (\ref{Q4}) respectively.
\koniec
\section{Eisenhart lift and Killing tensors}
\label{sec_eisenhart}
The Eisenhart metric in $(4+1)$ dimensions is 
\cite{Eisenhart}
\be
\label{eisenhart}
G=2 dudt +2 {V({\bf x}, t)}dt^2- d {\bf x}\cdot d{\bf x}.
\ee
The null geodesics of the metric $G$ satisfy
\[
\ddot{t}=0, \quad \dot{u}+2V\dot{t}=e, \quad \ddot{\bf x}=-\nabla V\dot{t}^2, \quad
\frac{1}{2}|\dot{\bf x}|^2+V=e,
\]
where $e$ is a constant. The first equation implies that $t$ can be used as a
parameter, with $\dot{t}=1$. The remaining equations then imply that null
geodesics of $G$ project to paths on the four--dimensional space of orbits of
the null isometry $\p/\p u$ which satisfy the Euler--Lagrange equations with the
potential $V$ and the energy given by $e$. Any first integral of degree $k$ of
the Euler--Lagrange equations on the space of orbits
\[
Q=Q_{i_1 i_2\cdots i_k}(x)\dot{x}^{i_1} \dot{x}^{i_2}\cdots \dot{x}^{i_k}+\dots
+ Q_i(x) \dot{x}^i+Q(x)
\]
lifts to a conformal Killing tensor of the metric (\ref{eisenhart}) represented
by a homogeneous function on $T\R^5$ given by
\[
\mathcal{Q}=Q_{i_1 i_2\cdots i_k}(x)\dot{x}^{i_1}\dot{x}^{i_2}\cdots \dot{x}^{i_k}+\dots+
Q_i(x)\dot{x}^i(\dot{t})^{k-1}+Q(x)(\dot{t})^k.
\]
The potential $V$ in (\ref{lagrangian}) gives rise to a family of Eisenhart metrics (\ref{Gtau}) parametrised by the frequency $\tau$. Therefore, the first integrals $Q_2$ given by (\ref{Q2}) and $Q_4$ given by (\ref{Q4}) give rise
to conformal Killing tensors (in fact, these are Killing tensors) of ranks two  and four 
for metrics  $G_2$ and $G_{1/2}$ respectively. Our results in \S{\ref{galois}} demonstrate that if 
$\tau^2\neq \{1, 4, 1/4\}$ then the Eisenhart  metric (\ref{Gtau}) does not admit an irreducible conformal Killing tensor such that the corresponding homogeneous function on  $T^*\R^{5}$ Poisson commutes with the $\phi$--momentum\footnote{The Eisenhart metric (\ref{eisenhart})
is Ricci--flat iff $V$ is a harmonic function. For the class of potentials (\ref{lagrangian}) this corresponds to $\tau^2=-2$.}. This establishes Corollary  \ref{color}.
\section{Projective metrisability of axi-symmetric potentials}
\label{sec_proj}
A three–dimensional path geometry on an open set
$U\subset\R^3$
is family of unparametrised
curves: one curve through any point of $U$ in any direction. Locally, a path geometry can
be represented by an equivalence class of systems of 2nd order ODEs
\be
\label{system}
\rho''=F(\phi,  \rho, z, \rho', z'), \quad z''=G(\phi, \rho, z, \rho', z')
\ee
where $(\rho, z)$ are functions of $\phi$, and $'=d/d\phi$. 
Two systems of the form (\ref{system}) are regarded as equivalent if they can be mapped to each-other by a diffeomorphism of $\R^3$. A path geometry arises from a {\em projective structure} if
there exists an affine connection $\nabla$ such that the geodesic
equations
\[
\ddot{x}^a+\Gamma^a_{bc}\dot{x}^b\dot{x}^c=0
\]
reduce to (\ref{system}) with $x^a=(\phi, \rho, z)$  after elimination of the affine parameter between the three equations,
and regarding $(x^2, x^3)=(\rho, z)$ as a function of 
$x^1=\phi$. If $\nabla$ is one such connection, then so is $\hat{\nabla}$ defined by the Christoffel symbols
\be
\label{projective}
  \hat{\Gamma}^a_{bc}=\Gamma^a_{bc}+\delta^a_b\Upsilon_c+\delta^a_c\Upsilon_b
\ee
where $\Upsilon=\Upsilon_bdx^b$ is any one--form. The projective structure $[\nabla]$ is then an equivalence class of connections defined by
(\ref{projective}). The necessary and sufficient conditions for the existence of projective structure for a system (\ref{system}) have been found by Fels \cite{fels}.

Let us consider a path geometry corresponding to 
the Euler--Lagrange equations
of a Lagrangian $L=T-V$  with an arbitrary axi-symmetric potential $V=V(\rho, z)$.  The generic initial data will have $h\equiv \rho^2\dot{\phi}\neq 0$. In this case
the integral curves of the EL equations can instead be parametrised by $\phi$, so
the  equations reduce to a system of second order ODEs of the form (\ref{system}). Using $\dot{z}=z'\dot{\phi}$ etc, where $'$ stands for $d/d\phi$ we find
this system to be
\begin{eqnarray}
\label{system1}
  \rho''&=&
  -\frac{\rho^4}{h^2}\frac{\p \Ve}{\p \rho}+\frac{2}{\rho}{(\rho')^2},\quad
  z''= -\frac{\rho^4}{h^2}\frac{\p \Ve}{\p z}+\frac{2}{\rho}{\rho' z'}\\
  &&\mbox{where}\quad \Ve=V+\frac{h^2}{2\rho^2}.\nonumber
\end{eqnarray}
The projective structure corresponding to 
(\ref{system1})
is readily found, and can be
represented by a connection $\nabla$  with Christoffel symbols
\be
\label{connectionV}
\Gamma_{\phi\rho}^{\phi}=\frac{3}{4\rho}, \quad
\Gamma_{\phi\phi}^{\rho}=\frac{\rho^4}{h^2}\frac{\p \Ve}{\p\rho}, \quad
\Gamma_{\rho\rho}^{\rho}=-\frac{1}{2\rho}, \quad  
\Gamma_{\phi\phi}^z=\frac{\rho^4}{h^2} \frac{\p \Ve}{\p z} , \quad
\Gamma_{\rho z}^z=-\frac{1}{4\rho}.
\ee
The question about the existence of an underlying metric is a more subtle one, and has
only been solved completely in dimension two \cite{BDE}. In dimension three some invariant
obstructions to metrisability have been found in 
\cite{DE, Emet}. We will now show that are they
sufficient to rule out metrizability for a potential given by (\ref{V1}).

%\be
%\nabla=-\frac{3}{4\rho}(d\phi\odot d\rho)\otimes\frac{\p}%{\p\phi}+
%\frac{\rho^4}{h^2}\frac{\p \Ve}{\p\rho} (d\phi\odot d\phi)%\otimes\frac{\p}{\p \rho}
%-\frac{1}{2 y}(d\rho\odot d\rho)\otimes \frac{\p}{\p \rho}
%+\frac{\rho^4}{h^2}(d\phi\odot d\phi)\frac{\p}{\p z}
%-\frac{1}{4\rho}(d\rho\odot dz)\otimes\frac{\p}{\p z}.
\noindent
{\bf Proof of Theorem \ref{theo2}}. The first step is to compute the curvature ${{R_{ab}}^c}_d$ of the connection (\ref{connectionV}). It is defined by
\[
[\nabla_a, \nabla_b]X^c={{R_{ab}}^c}_d X^d.
\]
The totally trace--free part of ${{R_{ab}}^c}_d$ is the projective Weyl tensor ${{W_{ab}}^c}_d$. This tensor is invariant under changes of a connection
(\ref{projective}) in a given projective structure.
Now define a traceless tensor ${\Phi^{ab}_c}$ (it was called 
$V$ in \cite{DE}) in terms of the projective curvature, and an arbitrary
section of $\Lambda^3{(\R^3)}$, which we chose to represent by totally antisymmetric tensor $\epsilon^{abc}$ with $\epsilon^{123}=1$
\[
\Phi^{ab}_c=\epsilon^{dea}W_{de}{}^b{}_c.
\]
It was shown in \cite{DE} that a necessary condition for metrisability is the existence of a non--degenerate
rank--two tensor $\sigma^{ab}$ such that
\be
\label{DEcond}
\Phi^{(ab}_d \sigma^{c)d}=0.
\ee
Setting $x^a=(\phi, \rho, z)$, and computing $\Phi$ from a connection (\ref{connectionV}) yields a tensor with components
$\Phi^{ab}_\rho=0, \Phi^{ab}_z=0$ and
\[
\Phi^{\rho\rho}_{\phi}=\frac{2\rho^4}{h^2}
\frac{\p^2 \Ve}{\p z\p\rho},\;
\Phi^{\rho z}_{\phi}=-\frac{\rho^3}{h^2}
\Big(\rho\frac{\p^2 \Ve}{\p \rho^2}-\rho\frac{\p^2\Ve}{\p z^2}
+3\frac{\p \Ve}{\p \rho} \Big), \;
\Phi^{zz}_{\phi}=-\frac{2\rho^3}{h^2}
\Big(\rho\frac{\p^2 \Ve}{\p\rho\p z}
+3\frac{\p \Ve}{\p z}\Big).
\]
%\[
%\Phi=\frac{2y^3}{h^2}\Big(2\frac{\p \Ve}{\p z\p {\rho}}  %\p_{\rho}\odot\p_{\rho} -
%2(\rho\frac{\p \Ve}{ \p{\rho}}-\frac{\p \Ve}{\p z^2}+3\frac{\p \Ve}{\p{\rho}})\p_{\rho}\odot\p_z -
%2(\rho \frac{\p \Ve}{\p {\rho}\p z}+3\frac{\p \Ve}{\p z})\p_z\odot\p_z \Big)\otimes d\phi.
%\]
From here we find that the projective curvature vanishes
if $\Ve=\mbox{const}/\rho^2$, so effectively $V=0$ as the constant
can be reabsorbed into $h$ in $\Ve$. Substituting $V$ as in (\ref{lagrangian}) in $\Ve$ we find that
the condition (\ref{DEcond}) implies $\sigma^{c\phi}=0$ for all~$c$. Therefore, $\sigma^{ab}$ is degenerate, and this projective structure is  not 
metrisable. We conclude that the potential $V$ in (\ref{lagrangian}) does not give rise to a metrisable path geometry.
\koniec

\end{document}